\documentclass[aps,prl,twocolumn,showpacs,superscriptaddress,groupedaddress]{revtex4}  
\usepackage{graphicx}  
\usepackage{dcolumn}   
\usepackage{bm}        
\usepackage{amssymb}   
\usepackage{sidecap}   
\usepackage{color}

\newcommand{\ppbar}{\mbox{$p\overline{p}$}}
\newcommand{\bbbar}{\mbox{$b \overline{b} $}}
\newcommand{\ttbar}{\mbox{$t \overline{t} $}}
\newcommand{\Pet}{\mbox{\ensuremath{\not \!\! P_T}}}
\newcommand{\met}{\mbox{\ensuremath{\not \!\! E_T}}}
\newcommand{\metx}{\mbox{\ensuremath{\not \!\! E_x}}}
\newcommand{\mety}{\mbox{\ensuremath{\not \!\! E_y}}}
\def \mvis {\ensuremath{ M_{\rm{vis}} }}

\begin{document}
\hspace{5.2in} \mbox{Fermilab-Pub-11-288-E}

\title{Search for Higgs bosons decaying to $\boldsymbol{\tau^+\tau^-}$ pairs in $\boldsymbol{p\overline{p}}$ collisions at $\boldsymbol{\sqrt{s}=1.96}$~TeV}

\affiliation{Universidad de Buenos Aires, Buenos Aires, Argentina}
\affiliation{LAFEX, Centro Brasileiro de Pesquisas F{\'\i}sicas, Rio de Janeiro, Brazil}
\affiliation{Universidade do Estado do Rio de Janeiro, Rio de Janeiro, Brazil}
\affiliation{Universidade Federal do ABC, Santo Andr\'e, Brazil}
\affiliation{Instituto de F\'{\i}sica Te\'orica, Universidade Estadual Paulista, S\~ao Paulo, Brazil}
\affiliation{Simon Fraser University, Vancouver, British Columbia, and York University, Toronto, Ontario, Canada}
\affiliation{University of Science and Technology of China, Hefei, People's Republic of China}
\affiliation{Universidad de los Andes, Bogot\'{a}, Colombia}
\affiliation{Charles University, Faculty of Mathematics and Physics, Center for Particle Physics, Prague, Czech Republic}
\affiliation{Czech Technical University in Prague, Prague, Czech Republic}
\affiliation{Center for Particle Physics, Institute of Physics, Academy of Sciences of the Czech Republic, Prague, Czech Republic}
\affiliation{Universidad San Francisco de Quito, Quito, Ecuador}
\affiliation{LPC, Universit\'e Blaise Pascal, CNRS/IN2P3, Clermont, France}
\affiliation{LPSC, Universit\'e Joseph Fourier Grenoble 1, CNRS/IN2P3, Institut National Polytechnique de Grenoble, Grenoble, France}
\affiliation{CPPM, Aix-Marseille Universit\'e, CNRS/IN2P3, Marseille, France}
\affiliation{LAL, Universit\'e Paris-Sud, CNRS/IN2P3, Orsay, France}
\affiliation{LPNHE, Universit\'es Paris VI and VII, CNRS/IN2P3, Paris, France}
\affiliation{CEA, Irfu, SPP, Saclay, France}
\affiliation{IPHC, Universit\'e de Strasbourg, CNRS/IN2P3, Strasbourg, France}
\affiliation{IPNL, Universit\'e Lyon 1, CNRS/IN2P3, Villeurbanne, France and Universit\'e de Lyon, Lyon, France}
\affiliation{III. Physikalisches Institut A, RWTH Aachen University, Aachen, Germany}
\affiliation{Physikalisches Institut, Universit{\"a}t Freiburg, Freiburg, Germany}
\affiliation{II. Physikalisches Institut, Georg-August-Universit{\"a}t G\"ottingen, G\"ottingen, Germany}
\affiliation{Institut f{\"u}r Physik, Universit{\"a}t Mainz, Mainz, Germany}
\affiliation{Ludwig-Maximilians-Universit{\"a}t M{\"u}nchen, M{\"u}nchen, Germany}
\affiliation{Fachbereich Physik, Bergische Universit{\"a}t Wuppertal, Wuppertal, Germany}
\affiliation{Panjab University, Chandigarh, India}
\affiliation{Delhi University, Delhi, India}
\affiliation{Tata Institute of Fundamental Research, Mumbai, India}
\affiliation{University College Dublin, Dublin, Ireland}
\affiliation{Korea Detector Laboratory, Korea University, Seoul, Korea}
\affiliation{CINVESTAV, Mexico City, Mexico}
\affiliation{Nikhef, Science Park, Amsterdam, the Netherlands}
\affiliation{Radboud University Nijmegen, Nijmegen, the Netherlands and Nikhef, Science Park, Amsterdam, the Netherlands}
\affiliation{Joint Institute for Nuclear Research, Dubna, Russia}
\affiliation{Institute for Theoretical and Experimental Physics, Moscow, Russia}
\affiliation{Moscow State University, Moscow, Russia}
\affiliation{Institute for High Energy Physics, Protvino, Russia}
\affiliation{Petersburg Nuclear Physics Institute, St. Petersburg, Russia}
\affiliation{Instituci\'{o} Catalana de Recerca i Estudis Avan\c{c}ats (ICREA) and Institut de F\'{i}sica d'Altes Energies (IFAE), Barcelona, Spain}
\affiliation{Stockholm University, Stockholm and Uppsala University, Uppsala, Sweden}
\affiliation{Lancaster University, Lancaster LA1 4YB, United Kingdom}
\affiliation{Imperial College London, London SW7 2AZ, United Kingdom}
\affiliation{The University of Manchester, Manchester M13 9PL, United Kingdom}
\affiliation{University of Arizona, Tucson, Arizona 85721, USA}
\affiliation{University of California Riverside, Riverside, California 92521, USA}
\affiliation{Florida State University, Tallahassee, Florida 32306, USA}
\affiliation{Fermi National Accelerator Laboratory, Batavia, Illinois 60510, USA}
\affiliation{University of Illinois at Chicago, Chicago, Illinois 60607, USA}
\affiliation{Northern Illinois University, DeKalb, Illinois 60115, USA}
\affiliation{Northwestern University, Evanston, Illinois 60208, USA}
\affiliation{Indiana University, Bloomington, Indiana 47405, USA}
\affiliation{Purdue University Calumet, Hammond, Indiana 46323, USA}
\affiliation{University of Notre Dame, Notre Dame, Indiana 46556, USA}
\affiliation{Iowa State University, Ames, Iowa 50011, USA}
\affiliation{University of Kansas, Lawrence, Kansas 66045, USA}
\affiliation{Kansas State University, Manhattan, Kansas 66506, USA}
\affiliation{Louisiana Tech University, Ruston, Louisiana 71272, USA}
\affiliation{Boston University, Boston, Massachusetts 02215, USA}
\affiliation{Northeastern University, Boston, Massachusetts 02115, USA}
\affiliation{University of Michigan, Ann Arbor, Michigan 48109, USA}
\affiliation{Michigan State University, East Lansing, Michigan 48824, USA}
\affiliation{University of Mississippi, University, Mississippi 38677, USA}
\affiliation{University of Nebraska, Lincoln, Nebraska 68588, USA}
\affiliation{Rutgers University, Piscataway, New Jersey 08855, USA}
\affiliation{Princeton University, Princeton, New Jersey 08544, USA}
\affiliation{State University of New York, Buffalo, New York 14260, USA}
\affiliation{Columbia University, New York, New York 10027, USA}
\affiliation{University of Rochester, Rochester, New York 14627, USA}
\affiliation{State University of New York, Stony Brook, New York 11794, USA}
\affiliation{Brookhaven National Laboratory, Upton, New York 11973, USA}
\affiliation{Langston University, Langston, Oklahoma 73050, USA}
\affiliation{University of Oklahoma, Norman, Oklahoma 73019, USA}
\affiliation{Oklahoma State University, Stillwater, Oklahoma 74078, USA}
\affiliation{Brown University, Providence, Rhode Island 02912, USA}
\affiliation{University of Texas, Arlington, Texas 76019, USA}
\affiliation{Southern Methodist University, Dallas, Texas 75275, USA}
\affiliation{Rice University, Houston, Texas 77005, USA}
\affiliation{University of Virginia, Charlottesville, Virginia 22901, USA}
\affiliation{University of Washington, Seattle, Washington 98195, USA}
\author{V.M.~Abazov} \affiliation{Joint Institute for Nuclear Research, Dubna, Russia}
\author{B.~Abbott} \affiliation{University of Oklahoma, Norman, Oklahoma 73019, USA}
\author{B.S.~Acharya} \affiliation{Tata Institute of Fundamental Research, Mumbai, India}
\author{M.~Adams} \affiliation{University of Illinois at Chicago, Chicago, Illinois 60607, USA}
\author{T.~Adams} \affiliation{Florida State University, Tallahassee, Florida 32306, USA}
\author{G.D.~Alexeev} \affiliation{Joint Institute for Nuclear Research, Dubna, Russia}
\author{G.~Alkhazov} \affiliation{Petersburg Nuclear Physics Institute, St. Petersburg, Russia}
\author{A.~Alton$^{a}$} \affiliation{University of Michigan, Ann Arbor, Michigan 48109, USA}
\author{G.~Alverson} \affiliation{Northeastern University, Boston, Massachusetts 02115, USA}
\author{G.A.~Alves} \affiliation{LAFEX, Centro Brasileiro de Pesquisas F{\'\i}sicas, Rio de Janeiro, Brazil}
% ...
%
%
\collaboration{The D0 Collaboration\footnote{with visitors from
$^{a}$Augustana College, Sioux Falls, SD, USA,
$^{b}$The University of Liverpool, Liverpool, UK,
$^{c}$SLAC, Menlo Park, CA, USA,
$^{d}$University College London, London, UK,
$^{e}$Centro de Investigacion en Computacion - IPN, Mexico City, Mexico,
$^{f}$ECFM, Universidad Autonoma de Sinaloa, Culiac\'an, Mexico,
and 
$^{g}$Universit{\"a}t Bern, Bern, Switzerland.
$^{\ddag}$Deceased.
}} \noaffiliation
\vskip 0.25cm
    
\date{\today}

\begin{abstract}
We present a search for the production of neutral Higgs bosons decaying 
into $\tau^{+}\tau^{-}$ pairs in $p\bar{p}$ collisions at a center-of-mass energy of 
$1.96$~TeV.   
The data, corresponding to an integrated luminosity of $5.4$~fb$^{-1}$,
were collected by the D0 experiment at the Fermilab Tevatron Collider.   
We set upper limits at the $95\%$~C.L.  
on the product of production cross section and branching ratio for a scalar
resonance decaying into $\tau^{+}\tau^{-}$ pairs, and we interpret these limits as limits on the production
of Higgs bosons in the minimal supersymmetric standard model (MSSM) and
as constraints in the MSSM parameter space.
\end{abstract}

\maketitle

Supersymmetry (SUSY)~\cite{susy} is one of the extensions
of the standard model (SM) proposed to address
its shortcomings, such as the hierarchy problem caused
by the quadratically divergent radiative corrections to the Higgs boson mass.
In the minimal supersymmetric standard model (MSSM),  
two complex Higgs boson doublets  lead to five physical Higgs bosons:  two neutral 
CP-even ($h$, $H$),  one neutral CP-odd ($A$), and two charged Higgs 
bosons ($H^{\pm}$).  The three neutral Higgs bosons ($h,H,A$) are
collectively denoted as $\phi$.
At tree level, the Higgs sector of the MSSM is fully described by two parameters, which 
are commonly chosen to be the mass of the 
CP-odd Higgs boson, $M_{A}$, and the ratio of the 
vacuum expectation values of the two Higgs doublets, $\tan \beta$.  
Radiative corrections introduce dependencies on additional MSSM parameters.   
The neutral MSSM Higgs bosons decay into $\tau^+\tau^-$ and $\bbbar$ pairs with branching 
ratios of $\approx 10\%$ and $\approx 90\%$, respectively.
Their production cross section is 
enhanced by a factor that depends on $\tan\beta$ with respect to the
cross section for the SM Higgs boson at the same Higgs boson mass.
Moreover, for large $\tan\beta$, the Higgs bosons $A$ and either $h$ or $H$
are  nearly degenerate in mass which leads to an effective doubling of
$\sigma_{\phi}(M_{\phi})$.

Searches for the production of
neutral MSSM Higgs bosons have been performed 
at the CERN $e^+e^-$ Collider (LEP)~\cite{lep}. 
The CDF and D0 Collaborations at the Fermilab Tevatron Collider
and the CMS Collaboration at the CERN Large Hadron Collider have excluded $M_A$ of up to $300$~GeV in 
a restricted region of $\tan\beta\approx 30-100$,
by searching for the exclusive processes
$(b)b\phi\to (b)bb\bar{b}$~\cite{tev-bbb} and
$b\phi\to b\tau^+\tau^-$~\cite{tev-btautau},
and for the inclusive process
$\phi\to\tau^+\tau^-$~\cite{tev-tautau,cdf-tautau,d0-tautau,cms-tautau}.

This Letter presents a search for the inclusive process
$gg,\bbbar\to\phi \rightarrow \tau^{+}\tau^{-}$, where the tau lepton pairs 
are reconstructed through their
decay into $e\mu$ or $\mu\tau_{h}$ final states, and $\tau_{h}$ represents the hadronic decay modes of the tau lepton.  The search for
$\tau^+\tau^-$ final states is interpreted
in a model-independent way and in the context of the MSSM.
The data were recorded with the D0 detector~\cite{d0det}
at a $p\bar{p}$ center-of-mass energy of $\sqrt{s}=1.96$~TeV and
correspond to an integrated luminosity of $5.4$~fb$^{-1}$. This represents
a significant increase compared to the results previously published by the 
CDF and D0 Collaborations, which
are based on integrated luminosities of $1.8$~fb$^{-1}$~\cite{cdf-tautau} and
$1.0$~fb$^{-1}$~\cite{d0-tautau}, respectively.

Signal samples are generated using the 
{\sc pythia}~\cite{pythia} Monte Carlo (MC) event generator with 
the CTEQ6L1 parton distribution functions (PDF)~\cite{cteq}.  
Dominant background processes comprise
$Z$+jets, $W$+jets, and multijet production.   
Background from multijet events arises when  
jets are misidentified as leptons.  Additional backgrounds include $t\bar{t}$ and
SM diboson production.   The backgrounds from $Z$+jets, $W$+jets, and $t\bar{t}$ 
production are modeled using {\sc alpgen}~\cite{bib-alpgen},
with parton showering and hadronization provided by {\sc pythia}.   
The {\sc alpgen}-generated samples make use of the MLM~\cite{bib-mlm} jet-parton
matching scheme to improve the jet multiplicity modeling. 
Diboson processes ({\sl WW, WZ, ZZ}) are simulated using {\sc pythia}.   
In all cases {\sc tauola}~\cite{tauola} is used to model the tau lepton decays. 
Simulated events are then processed by a {\sc geant}-based~\cite{geant} 
simulation of the D0 detector, and data events from random 
beam crossings are overlaid to model detector noise and multiple $\ppbar$ interactions.
Higher order quantum chromodynamics (QCD) calculations of cross sections are 
used to normalize the simulated background samples, except for
the background from multijet production, for which
the normalization and differential distributions are derived from data.

Events are selected by requiring at least one single muon 
trigger for the $\mu\tau_{h}$ channel, while for the $e\mu$ channel, 
they need to fulfill either
inclusive electron or muon trigger conditions.
Electrons are reconstructed using their characteristic energy deposits,  
including the transverse and longitudinal shower profiles in the
electromagnetic (EM) calorimeter.   
Muons are identified
by combining tracks in the central tracking detector with patterns of hits in the muon spectrometer.
Electrons and muons are required to be isolated in the calorimeter and
in the tracking detectors.
 
Tau lepton decays into hadrons are characterized as narrow, isolated 
jets with lower track multiplicity than quark or gluon jets.
Three types of tau lepton decays are distinguished by their detector signature.
One-prong tau decays consisting of energy deposited primarily in the hadronic calorimeter 
associated with a single track ($\pi^\pm\nu$-like) are denoted as $\tau$-type~1;
$\tau$-type~2 corresponds to one-prong tau decays with energy deposited in both
the hadronic and EM
calorimeters, associated with a single track ($\rho^\pm$$\nu$-like); and
$\tau$-type~3 are multi-prong decays with energy 
in the calorimeter and two or more associated tracks 
with invariant mass below $1.7$~GeV. 
A calibration for the energy of $\tau_h$ candidates measured in the calorimeter is  
derived from data.
It is based on the ratio of the calorimeter energy 
and the transverse momentum, $p_{T}$, measured in the tracking detector
for the $\tau_h$ candidates. The ratio is adjusted
in the simulation to match the data as a function
of the fraction of the $\tau_h$ energy deposited in the EM calorimeter.
  
A set of neural networks, one for each $\tau$-type, is applied to 
discriminate hadronic tau decays from jets~\cite{d0-z-tautau}.   
The input variables are related to isolation and shower shapes, 
and exploit correlations between calorimeter energy deposits and tracks.  
When requiring the neural network discriminants ({\sl NN}$_{\tau}$) to
be {\sl NN}$_{\tau}>0.9$ for $\tau$-types~1, 2
and {\sl NN}$_{\tau}>0.95$ for $\tau$-type~3, 
approximately $67\%$ of $Z/\gamma^*\rightarrow\tau^+\tau^-$~events
are retained in data, while $98\%$~of the multijet background events are 
rejected.

A series of selections is used to reduce the background from $Z$+jets, $W$+jets, and multijet production.  The $Z/\gamma^{*} \rightarrow \tau^-\tau^+$
process differs from a Higgs boson signal only through the mass and spin
of the produced resonance and cannot be further reduced.
One isolated muon with $p_{T}^{\mu} >15$~GeV and an isolated hadronic tau lepton with 
transverse energy $E_{T}^{\tau} > 12.5$~GeV
($\tau$-types~$1,2$) or $E_{T}^{\tau} > 15$~GeV ($\tau$-type 3) 
are required in the $\mu\tau_{h}$ channel. 
The muon and the $\tau_h$ must be oppositely charged, where the charge
of the $\tau_h$ candidate is determined by the curvature of the associated track.
For $\tau$-type~$3$ the charge is obtained by summing over all tracks associated
with the $\tau_h$.
The pseudorapidity $\eta$~\cite{eta} is required to be $|\eta_{\mu}| <1.6$ 
for muons  and $|\eta_{\tau}| <2.5$ for tau leptons.   
The transverse momentum sums of all tracks associated with the
$\tau_h$ candidate, $p_{T}^{\tau}$, are required to be greater than 
$7$, $5$, $10$~GeV for $\tau$-types 1, 2, and 3, respectively.  At least one hit in the active 
layers of the D0 silicon vertex detector is required for the tracks associated with the
$\tau_h$.
The $\tau_{h}$ and the muon are required to originate from the same $\ppbar$ 
vertex and must  be separated 
from each other by $\Delta{\cal R}=\sqrt{(\Delta\eta)^{2}+(\Delta\varphi)^{2}} > 0.5$, where
$\Delta\varphi$ is the difference in azimuthal angle. 
This requirement suppresses the 
$Z/\gamma^{*} \rightarrow \mu^+\mu^-$ background.   
The transverse $W$ boson mass in $W\to\ell\nu$ events is given by
$M_T^{\ell\nu}=\sqrt{2p^{\ell}_{T}\met \left[ 1-\cos(\Delta\varphi(\ell,\met) \right] }$ with $\ell=e,\mu$.
The components $\metx$ and $\mety$ of the missing transverse energy, $\met$,
are computed from calorimeter cells and the momenta of muons, and corrected for the energy 
response of electrons, tau leptons, and jets.   
We require $M_T^{\mu\nu}<50$~GeV to reject $W(\to \mu \nu)$+jets events where 
jets are misidentified as $\tau_h$ candidates. 

In the $e\mu$ channel, events with at least one muon with $p_T^{\mu} >10$~GeV and
$|\eta_\mu|<1.6$,
and an oppositely charged electron with $p_T^{e} > 12$~GeV and $|\eta_e|<2$
are selected. The $e\mu$ pair formed by the leptons with
the highest $p_T$ are selected as a candidate; they must be separated by $\Delta{\cal R} > 0.4$.  
To reject $Z\to\mu\mu\gamma$ events, an electron candidate is rejected
if it shares the same track with a muon.
Multijet background and $W$ boson production are suppressed by requiring the 
mass of the $e\mu$ pair to be larger than $20$~GeV and 
$\met + p_{T}^{\mu} + p_{T}^e > 65$~GeV.
Background from $W$+jets production is reduced by requiring 
$\min\left\{M_T^{e\nu},M_T^{\mu\nu}\right\}<10$~GeV.
The difference in the azimuthal angle, $\Delta\varphi(\ell,\met)$,
has to be $<0.3$ where $\ell=e,\mu$ is the lepton 
with the smaller $p_T$. This requirement
rejects background from {\it WW}, $t\bar{t}$, and $W$+jets production.   
Requiring the
scalar sum of the transverse momenta of all jets to be $<70$~GeV
rejects a large fraction of $t\bar{t}$ events.

To determine the expected background contribution from multijet production
in the  $\mu\tau_{h}$ channel, two {\sl NN}$_{\tau}$ regions
are selected in addition to the high {\sl NN}$_{\tau}$ ``signal'' region defined 
previously: the ``medium'' region in
the range $0.25 \leq$ {\sl NN}$_{\tau} \leq 0.75$ and the ``low" region
with {\sl NN}$_{\tau} \leq 0.1$. The samples are further divided depending
on whether the muon and the $\tau_h$ candidate have the same or 
opposite charge.  Background from $W$+jets production in these samples is reduced
by requiring $M_T^{\mu\nu}<50$~GeV. 
The transverse mass is calculated from the $\met$ and from the
azimuthal angle $\Delta\varphi(\mu,\met)$ between the direction of the muon transverse
momentum $p^{\mu}_{T}$ and the $\met$. 
The estimated contribution from MC-simulated background processes 
is then subtracted from the resulting distributions, and
the shape of the multijet background is derived from
the distributions of same-sign $\mu\tau_{h}$ pairs with {\sl NN}$_{\tau}> 0.9$.  
Multijet events mainly populate the low {\sl NN}$_{\tau}$ region, 
and the ratio of opposite to same-sign $\mu\tau_{h}$ pair events in this region
is extrapolated to yield the normalization of multijet events in the signal sample.   
This estimate of the multijet background contribution is verified by an 
independent method which uses the medium {\sl NN}$_{\tau}$ region.
The difference between the estimates obtained by the two methods is
used as systematic uncertainty on the multijet background.
  
Multijet background in the $e\mu$ channel is determined by applying the 
same selection criteria as for signal apart from the electron likelihood and muon isolation criteria, which are inverted.  
The normalization is then taken from the ratio of the numbers of events in 
the opposite and same-sign samples.

\begin{figure}[htb]
\begin{center}
\includegraphics[height=68mm]{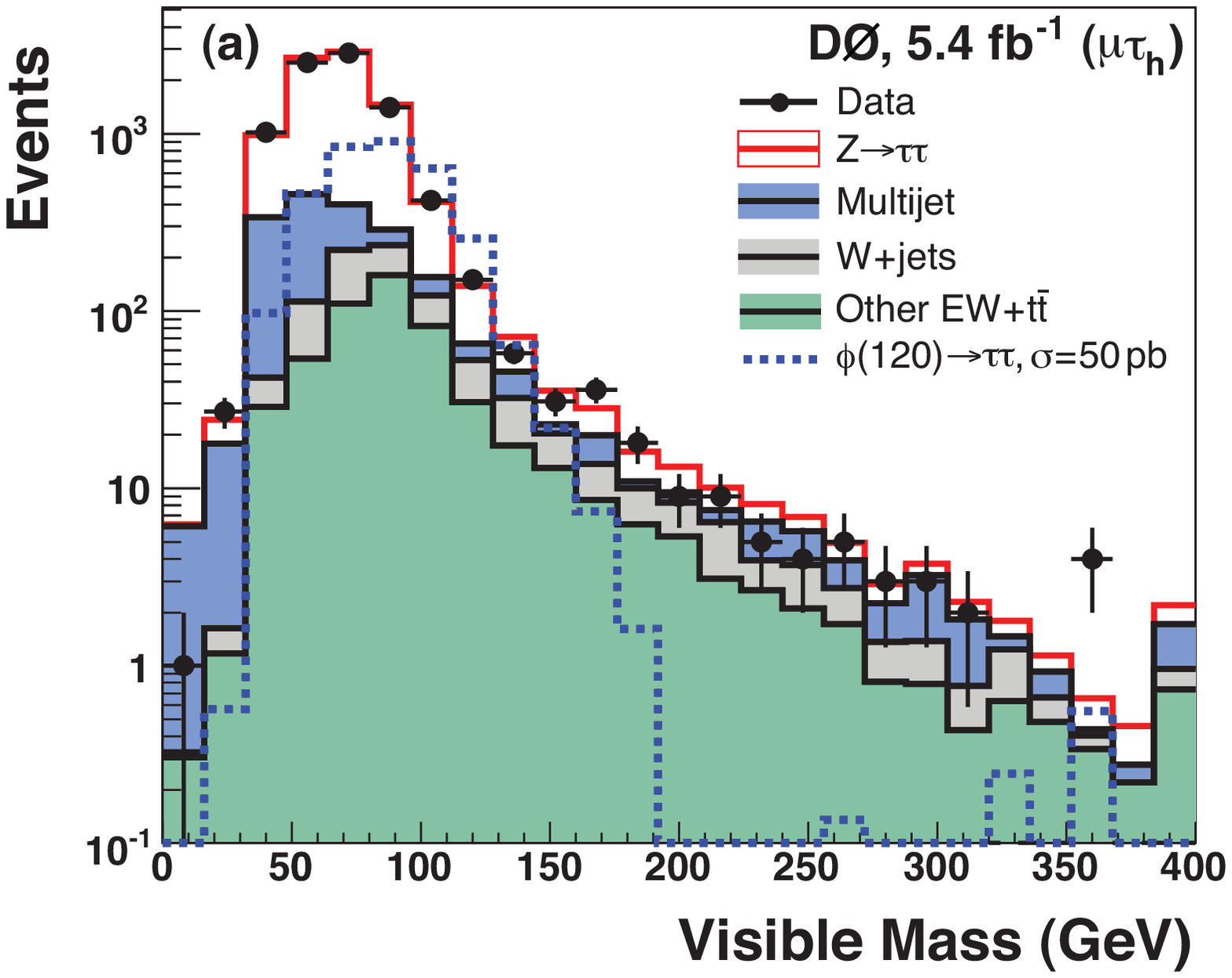}
\includegraphics[height=64mm]{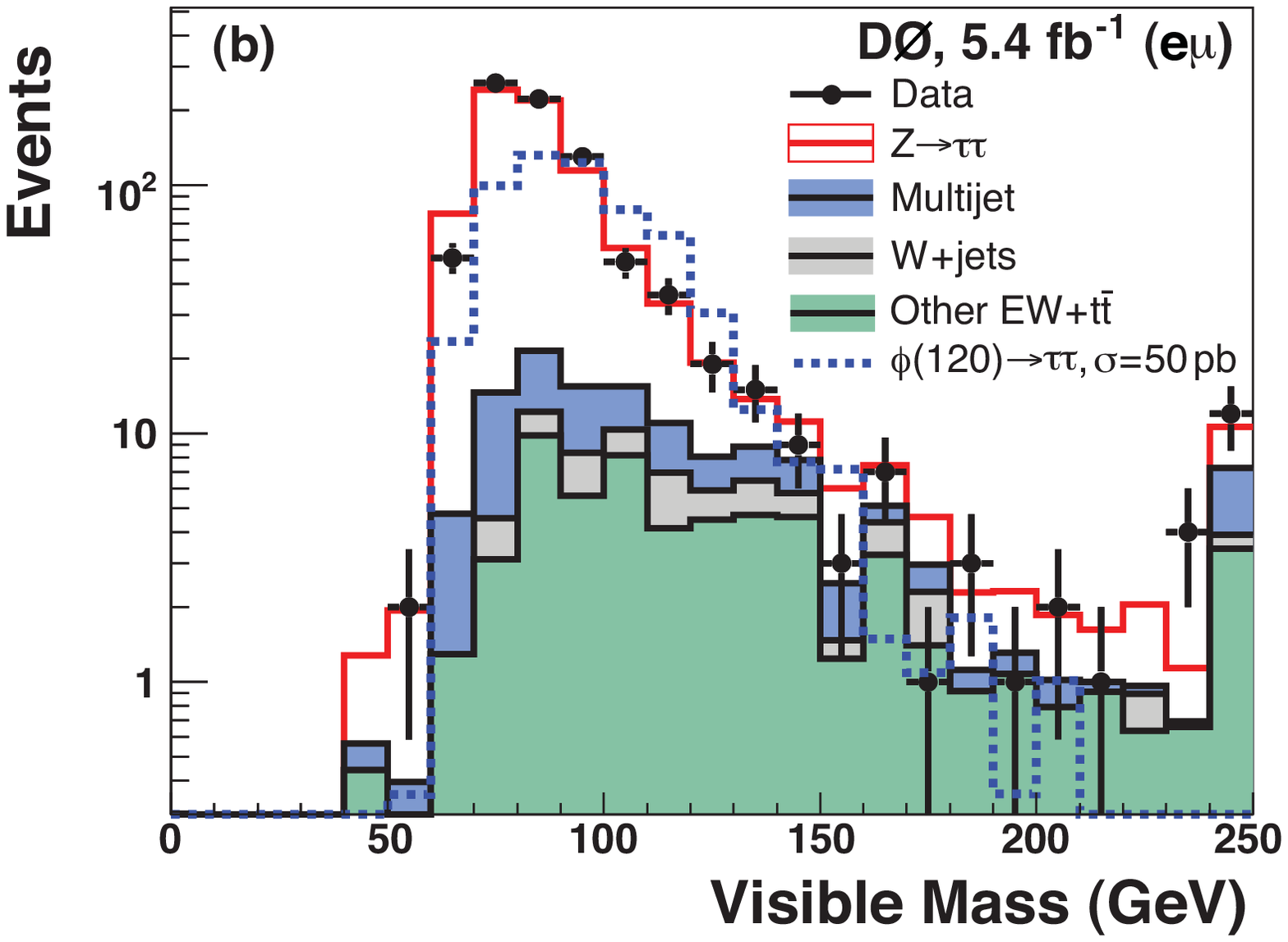}
\caption{Distributions of $\mvis$ in the (a) $\mu\tau_h$ and (b) $e\mu$ channels
after all selections.
The data, shown with statistical uncertainties, are compared to the sum of the predicted backgrounds for an integrated luminosity of 5.4~fb$^{-1}$.
The Higgs boson signal for $M_{\phi}=120$~GeV 
is normalized to a production cross section of 
$\sigma_{\phi}=50$~pb. 
All entries exceeding the range of a
     histogram are added to the last bin.
}\label{f:vis-mass}
\end{center}
\vspace{-3mm}
\end{figure}

We search for
an enhancement from a $\tau^+\tau^-$ resonance  above the expected background in the distribution
of the visible mass $\mvis=\sqrt{(P_{\tau_1}+  P_{\tau_2} +  \ \Pet)^2}$, 
which is calculated using the four-vectors of the measured tau lepton decay 
products, $P_{\tau_{1,2}}$, and the missing transverse momentum, 
$\Pet = (\met,\metx,\mety, 0)$. 
In the $e\mu$ final state, the four-vectors $P_{\tau_{1,2}}$ 
are calculated using the reconstructed electron and muon, respectively.
After imposing all selection requirements,  the $\mvis$ distributions for the $\mu\tau_{h}$ and $e\mu$ final states are 
shown in Fig.~\ref{f:vis-mass}.
Table~\ref{t:evt_numbers} gives the yields of the predicted background and of
data, summed over the $\mvis$ distributions shown in Fig.~\ref{f:vis-mass}.

\begin{table}
\caption{Expected number of events for backgrounds, number of events observed in data and 
efficiency, relative to all $\tau$ lepton decays,
for a signal with \mbox{$M_\phi=120$~GeV} summed over the $\mvis$ distributions shown
in Fig.~\protect\ref{f:vis-mass}. The total uncertainties are also given.
 \label{t:evt_numbers}}
\begin{tabular}{lcc}
\hline\hline
Channel         & ~~~~~~$\mu \tau_h$~~~~~~  & ~~~~~~$e \mu$~~~~~~ \\
\hline
$Z / \gamma^* \to \tau^+ \tau^-$     	&   $6914\pm591$  & $697\pm55$ \\
Multijet                         	        		&   $972\pm98$      & $53\pm8$  \\ 
$W \to e \nu, \mu \nu, \tau \nu$ 		&   $363\pm60$      & $19\pm5$    \\
$Z/\gamma^*\to e^+e^-,\mu^+\mu^-$&   $353\pm32$      & $34\pm 6$  \\
Diboson + $\ttbar$              		&   $180\pm12$      & $27\pm 5$  \\ \hline
Total Background                   		&   $8782\pm603$  & $830\pm56$  \\ \hline
Data                                                   &   $8574$              & $825$ \\  \hline
Efficiency (\%)                    		&   $1.16\pm 0.03$  &  $0.20\pm 0.01$ \\
\hline
\hline
\end{tabular}
\end{table}

Several sources of systematic
uncertainty affect both the signal efficiency and background
estimation.
Both uncertainties that modify only the normalization and uncertainties that 
change the shape of the $\mvis$ distribution are taken into account. 
Those that affect the 
normalization include the integrated luminosity ($6.1\%)$, 
muon identification efficiency ($2.9\%$), $\tau_h$ identification 
($12\%$, $4.2\%$,  and $7\%$ for $\tau$-types 1, 2, and 3, respectively), efficiency to reconstruct the
$\tau_h$ track ($1.4\%$), electron identification efficiency ($3.5\%$),
PDF uncertainty on the acceptance ($4.6\%$),  
the uncertainty on the $Z$+jets cross sections ($5\%$), the $W$+jets 
normalization ($10\%$ for $e\mu$ and $20\%$ for $\mu\tau_h$), $t\bar{t}$ cross section 
($10\%$), diboson cross section ($6\%$), 
muon and electron trigger efficiencies (both $5\%$), jet energy scale ($1.5\%-2\%$),
and the modeling of the multijet background ($9.1\%$, $17.7\%$, and $12.5\%$ 
for $\tau$-types 1, 2, and 3, respectively).   
Uncertainties arising from modeling of the $Z$ boson transverse momentum
and the $\tau_h$ energy scales modify the shape of the $\mvis$ distribution.

\begin{figure}[ht!!!]
\begin{center}
\includegraphics[height=70mm]{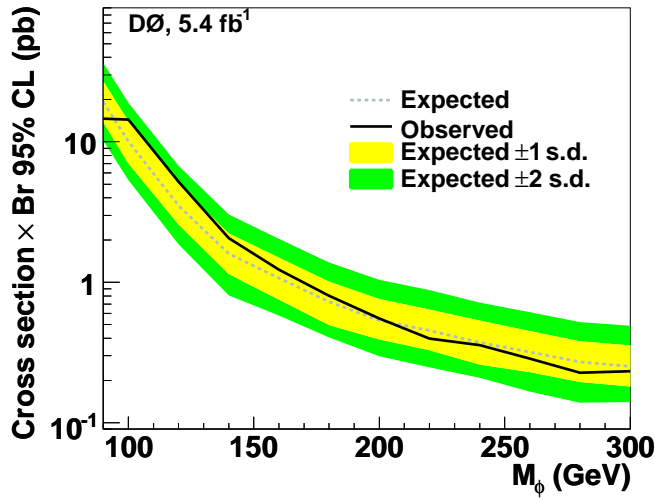}
\caption{
Model-independent expected and observed $95\%$~C.L. upper limits on the product of production cross section and branching ratio for
inclusive $\ppbar\to\phi \to \tau^+\tau^-$ production as a function of $M_{\phi}$, 
assuming a SM total width for the Higgs boson.
The $\pm 1,2$~standard deviation (s.d.)~variations of the expected limits are
shown as bands.
}
\label{f:xs-limit}
\end{center}
\vspace{-3mm}
\end{figure}

The $\mvis$ distribution is used to calculate 
upper limits on the cross section based on a modified 
frequentist method with a 
log-likelihood ratio test statistics~\cite{bib-junk}
and a profiling technique to reduce the impact of systematic 
uncertainties~\cite{bib-collie}.  The value of {\sl CL}$_s$, is calculated as 
{\sl CL}$_s =~${\sl CL}$_{s+b} / ${\sl CL}$_b$,
where {\sl CL}$_{s+b}$ and {\sl CL}$_b$~are the $p$-values under 
signal+background and background-only hypotheses, respectively.   
The expected and observed limits are calculated by scaling the signal until $1-\mbox{\sl CL}_s$ reaches 0.95. 
The combined limits on the 
product of production cross section and branching ratio into tau lepton pairs are given in Fig.~\ref{f:xs-limit} and Table~\ref{t:limits} as a function of $M_{\phi}$. 
The combined limits assume a scalar resonance with the decay width of a SM Higgs boson,
which is negligible compared to the experimental resolution on $\mvis$.  

\begin{table}
\caption{Upper limits on the 
expected and observed cross sections 
(in pb) multiplied by the branching ratio for $\phi\to\tau^+\tau^-$ at the $95\%$~C.L.~as a function of $M_{\phi}$ (in GeV).
\label{t:limits}}
\begin{tabular}{c|cccccc}
\hline\hline
$M_{\phi}$  & Observed  & $-1$~s.d.  &  Expected & $+1$~s.d. \\
\hline   
  90 & 14.7 &  13.8 & 19.2 & 27.1 \\
100 & 14.4 &  7.00 & 10.1 & 14.0 \\
120 & 5.22 &  2.58 & 3.53 & 5.01 \\
140 & 2.06 & 1.14  & 1.60 & 2.23 \\
160 & 1.23 & 0.75  & 1.07 & 1.50 \\
180 & 0.80 & 0.50  & 0.73 & 1.01 \\
200 & 0.55 & 0.39  & 0.54 & 0.76 \\
220 & 0.40 & 0.33  & 0.45 & 0.64 \\
240 & 0.36 & 0.26  & 0.37 & 0.53 \\
260 & 0.29 & 0.23  & 0.32 & 0.45 \\
280 & 0.23 & 0.19  & 0.27 & 0.38 \\
300 & 0.23 & 0.18  & 0.25 & 0.36 \\
\hline\hline
\end{tabular}
\end{table}

\begin{figure*}[htbp]
\begin{center}
\includegraphics[height=75mm]{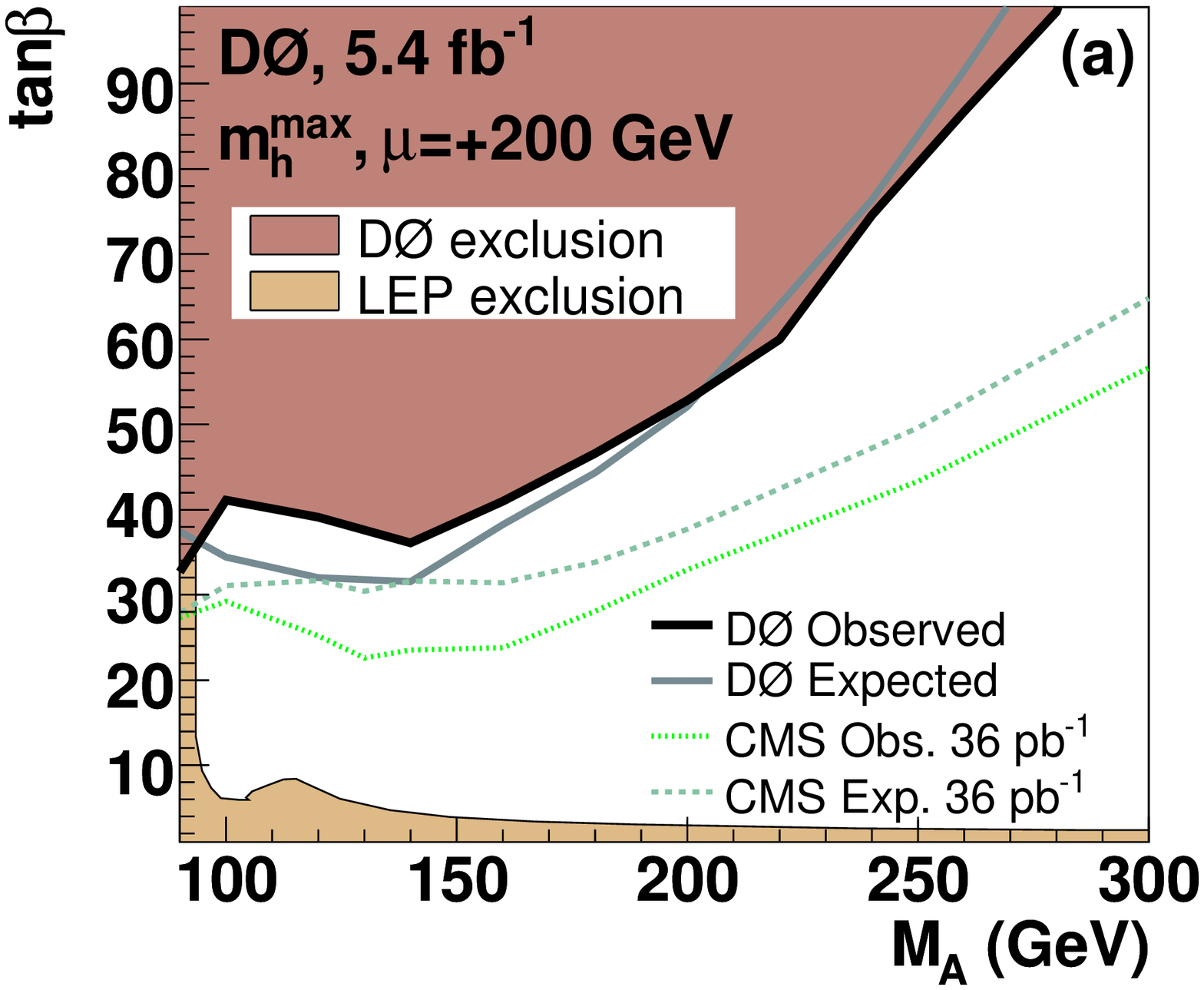}
\includegraphics[height=75mm]{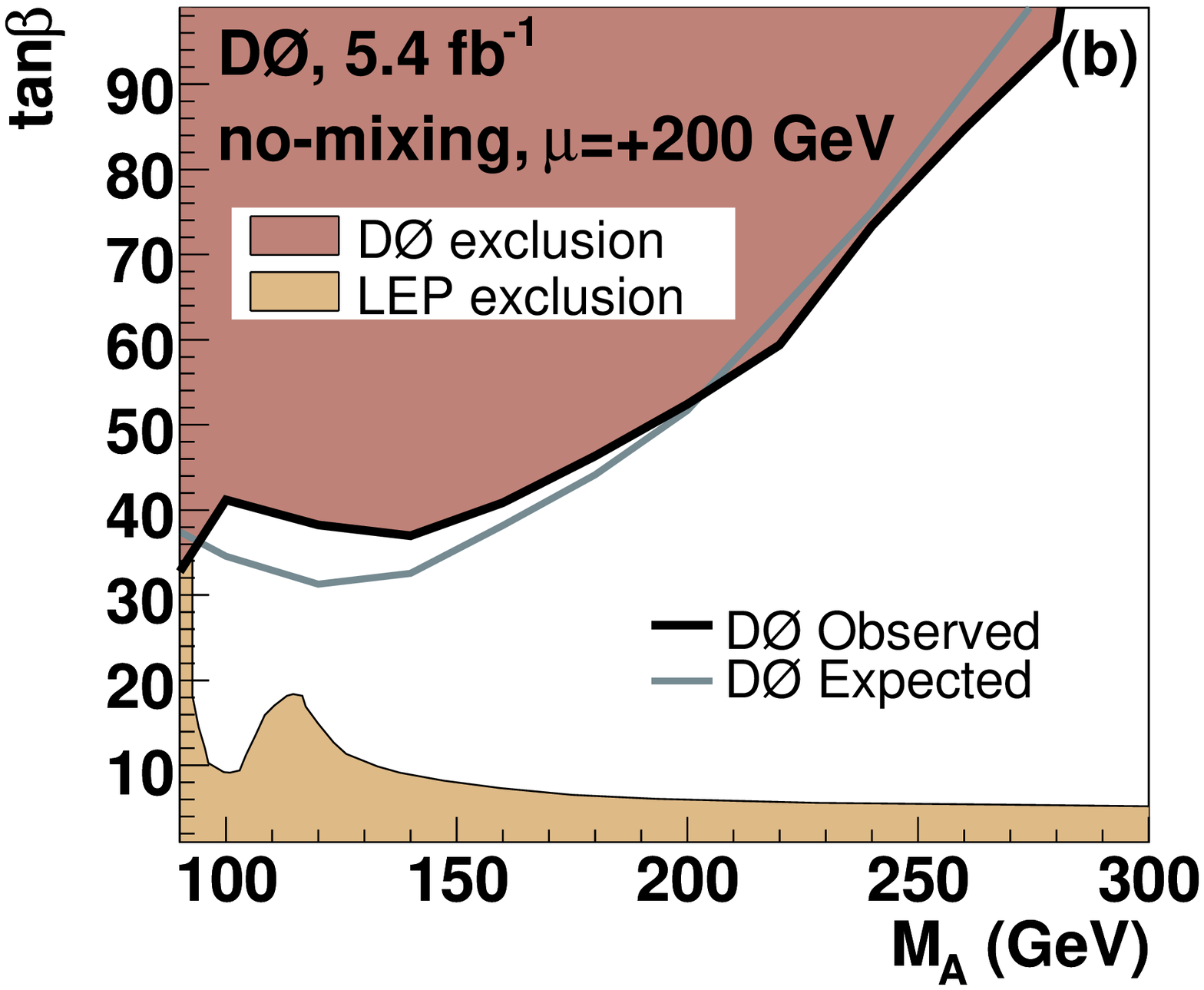}
\caption{
 Expected and observed exclusion regions at the $95\%$ C.L. in
 the plane of $\tan\beta$ versus $M_A$ for the (a) $m_h^{\rm max}$ and 
 (b) no-mixing scenarios with $\mu = +200$~GeV. 
 The regions excluded by the LEP Collaborations~\cite{lep}
 and the CMS Collaboration~\cite{cms-tautau} are also shown.}
\label{f:tanb-excl}
\end{center}
\vspace{-3mm}
\end{figure*}

In addition to $M_{A}$ and $\tan\beta$, the masses and couplings of the 
Higgs bosons in the MSSM depend on additional
parameters through radiative corrections.   
The production cross section limits are therefore translated into
exclusions in the tan$\beta$ versus $M_{A}$ plane for two representative MSSM scenarios assuming a CP-conserving Higgs 
sector~\cite{mssm-benchmark},
the $m_h^{\rm max}$ scenario~\cite{mhmax} and the 
no-mixing scenario~\cite{no-mixing} 
with a Higgs mass parameter $\mu=+200$~GeV.   
The signal cross sections and branching ratios are calculated using the
{\sc feynhiggs}~\cite{feynhiggs} program, where the $gg \to \phi$ production cross section is taken
from~\cite{theory} and the $b\bar{b} \to \phi$ production cross section from~\cite{bib-harl}.

At large values of tan$\beta$, the Higgs boson width 
increases with $\tan\beta$ and 
can become significantly larger than the value in the SM.  
This effect was previously studied by  convolving 
a relativistic Breit-Wigner
function with the next-to-leading order calculation of the signal cross section 
from {\sc feynhiggs} as a function of $M_{\phi}$ and $\tan\beta$~\cite{d0-tautau}. 
In the $(M_A,\tan\beta)$ region where this analysis sets $95\%$ C.L. limits,
and for  $\mu=+200$~GeV, the Higgs boson width is smaller than $0.1M_{\phi}$ and 
less than half of the experimental resolution on $\mvis$. 
The signal cross section in this channel is largely
insensitive to ${\rm sign}(\mu)$.
The ratio of the $gg\to\phi$ and $\bbbar\to\phi$ cross sections also depends on
$\tan\beta$. For this inclusive search, the difference between the efficiencies of
the two production mechanisms is small and can be neglected.

The region in the MSSM parameter space excluded at the $95\%$ C.L. is
shown in Fig.~\ref{f:tanb-excl} up to $M_A=300$~GeV.
For $M_A\approx 140$~GeV, the expected exclusion 
reaches $\tan\beta\approx 30$, which
is comparable to recent limits obtained in~\cite{cms-tautau}.
The upper limits on the  product of the $\ppbar$
production cross section for a neutral Higgs boson and branching ratio into tau leptons
represent the most stringent limits to date.

%
We thank the staffs at Fermilab and collaborating institutions,
and acknowledge support from the
DOE and NSF (USA);
CEA and CNRS/IN2P3 (France);
FASI, Rosatom and RFBR (Russia);
CNPq, FAPERJ, FAPESP and FUNDUNESP (Brazil);
DAE and DST (India);
Colciencias (Colombia);
CONACyT (Mexico);
KRF and KOSEF (Korea);
CONICET and UBACyT (Argentina);
FOM (The Netherlands);
STFC and the Royal Society (United Kingdom);
MSMT and GACR (Czech Republic);
CRC Program and NSERC (Canada);
BMBF and DFG (Germany);
SFI (Ireland);
The Swedish Research Council (Sweden);
and
CAS and CNSF (China).
%


\begin{thebibliography}{99}
\bibitem{eta}
The pseudorapidity is defined as $\eta = -\ln[\tan(\theta/2)]$, where $\theta$ is the polar angle with respect to the proton beam direction.
\end{thebibliography}
\end{document}